\documentclass[a4paper,pra,twocolumn,superscriptaddress]{revtex4-1}

\usepackage{amssymb}
\usepackage{amsmath}
\usepackage{epsfig}
\usepackage{color}
\usepackage{graphics, graphicx}
\usepackage{bbold}
\usepackage{psfrag}
\usepackage{mathcomp}
\usepackage{subfigure}
\usepackage{verbatim}
\usepackage{color}
\usepackage{blindtext}
\usepackage{float}
\usepackage[utf8]{inputenc}
\usepackage[T1]{fontenc}
\usepackage{supertabular}
\usepackage{longtable}
\usepackage[colorlinks,citecolor=blue]{hyperref}

\begin{document}

\date{\today}
\title{Nonadiabatic transitions in non-Hermitian $\mathcal{PT}$-symmetric two-level systems}

\author{Jian-Song Pan}
\email{panjsong@scu.edu.cn}
\affiliation{College of Physics, Sichuan University, Chengdu 610065, China}
\affiliation{Key Laboratory of High Energy Density Physics and Technology of Ministry of Education, Sichuan University, Chengdu 610065, China}

\author{Fan Wu}
\email{T21060@fzu.edu.cn}
\affiliation{Fujian Key Laboratory of Quantum Information and Quantum\\
	Optics, College of Physics and Information Engineering, Fuzhou University,
	Fuzhou, Fujian, 350108, China\\}

\begin{abstract}
{We systematically characterize the dynamical evolution of time-parity ($\mathcal{PT}$)-symmetric two-level systems with spin-dependent dissipations. If the control parameters of the gap are linearly tuned with time, the dynamical evolution can be characterized with parabolic cylinder equations which can be analytically solved. We find that the asymptotic behaviors of particle probability on the two levels show initial-state-independent redistribution in the slow-tuning-speed limit as long as the system is nonadiabatically driven across exceptional points. Equal distributions appear when the nondissipative Hamiltonian shows gap closing. So long as the nondissipative Hamiltonian displays level anticrossing, the final distribution becomes unbalanced. The ratios between the occupation probabilities are given analytically. These results are confirmed with numerical simulations. The predicted equal distribution phenomenon may be used to identify the closing of the energy gap from anti-crossing between two energy bands.}
\end{abstract}

\maketitle
\section{Introduction}
The past decades have witnessed the rapid growth of research interest in non-Hermitian systems with parity-time ($\mathcal{PT}$) symmetry~\cite{bender1998real,el2018non,bender2018pt,moiseyev2011non,feng2017non,pile2017gaining, limonov2017fano, horiuchi2017marriage,xiao2019observation,klauck2019observation,szameit2011p,regensburger2012parity,zhen2015spawning,weimann2017topologically, kim2016direct,cerjan2016exceptional,poshakinskiy2016multiple,dembowski2001experimental,dembowski2004encircling,mailybaev2005geometric,doppler2016dynamically,hassan2017chiral,hassan2017dynamically,
zhang2018dynamically,ren2022chiral}. These systems possess real (conjugate complex) spectra in the $\mathcal{PT}$-symmetric (breaking) phases~\cite{bender1998real,el2018non,bender2018pt}. The spontaneous break of $\mathcal{PT}$ symmetry by spin-dependent dissipation (or gain and loss in optics) leads to the emergence of an exceptional point (EP) with the coalescence of eigenstates, which has been experimentally checked on different platforms~\cite{moiseyev2011non,feng2017non,pile2017gaining, limonov2017fano, horiuchi2017marriage,xiao2019observation,klauck2019observation,szameit2011p,regensburger2012parity,zhen2015spawning,weimann2017topologically, kim2016direct,cerjan2016exceptional,poshakinskiy2016multiple}. The singular character of EPs is not only useful for sensing~\cite{wiersig2016sensors,wiersig2014enhancing,wiersig2020review}, but has a profound impact on the dynamics of the system, exemplified by
the phenomenon of chiral state transfer in the dynamical evolution surrounding EPs~\cite{dembowski2004encircling,hassan2017dynamically,hassan2017chiral,zhang2018dynamically,wang2018non,ren2022chiral}.

{Furthermore, when a $\mathcal{PT}$ symmetric two-level system is driven directly through EPs, the transition probability shows anomalous asymptotic behaviours~\cite{longstaff2019nonadiabatic}. In the adiabatic limit, an equal redistribution between the states coalescing at the exceptional points is observed. Equal redistribution is numerically shown to be independent of the initial states. However, only a typical $\mathcal{PT}$ symmetric model $H=\eta \sigma_{z}+i\gamma\sigma_{x}$ has been taken into account. In fact, a generic $\mathcal{PT}$-symmetric model should take the form of $H=\eta \sigma_{z}+\delta_{0}\sigma_{0}+\delta_{y}\sigma_{y}+i\gamma\sigma_{x}$~\cite{okugawa2019topological}. The natural question then is whether these phenomena are preserved for the generic $\mathcal{PT}$-symmetric two-level model. In addition, if the loss of initial information can be analytically proven, and how the additional real $\sigma_{y}$ term influences the asymptotic redistribution. To answer these questions, a systematic study on the nonadiabatic transition of the generic $\mathcal{PT}$-symmetric model is required.}

{In this work, we systematically characterize the nonadiabatic time evolution of the generic $\mathcal{PT}$-symmetric non-Hermitian two-level model with spin-dependent dissipations. The nondissipative Hamiltonian shows gap closing or anti-crossing when it possesses $\mathcal{PT}$ symmetry and shows level anti-crossing when the $\mathcal{PT}$ symmetry is absent. A section of imaginary spectra ended with EPs ($\mathcal{PT}$-symmetry-breaking bubble) emerges, provided that the $\mathcal{PT}$ symmetry is broken and retrieved by tuning the gap-control parameter. When the energy gap control parameter is tuned to cross the EPs, the particle probabilities of nonadiabatic evolution are shown to be redistributed on the two levels in the slow-tuning-speed limit. The probability ratio between the two levels is a constant determined by the Hamiltonian parameters, which is independent of the initial states.

We find that the asymptotic behavior of equal distribution only exists when the nondissipative Hamiltonian shows the closure of the gap, that is, the case studied in the previous work~\cite{longstaff2019nonadiabatic}. The redistribution becomes unbalanced if the nondissipative Hamiltonian displays level anti-crossing, i.e., the currently focused case with the generic $\mathcal{PT}$-symmetric model. These analytical results are confirmed in numerical simulations. For comparison, cases without $\mathcal{PT}$ symmetries are also addressed, where the initial-state-independent asymptotic behavior disappears. It is worth to emphasize that the redistribution asymptotic behavior is in sharp contrast to the Landau-Zener-Stückelberg interference in a common Hermitian system, where the final state sensitively depends on initial condition.}

Many interesting physical processes, such as first-order quantum phase transitions, are signaled by closing the energy gap between the ground and the first excited states~\cite{sachdev2007quantum}. Based on the unique dynamic consequence of EPs discovered here, we propose to detect the gap-closing transition of the nondissipative system through a Landau-Zener-Stückelberg-like process: the system parameters (i.e., the quasimomentum driven by static force for an energy band~\cite{longstaff2019nonadiabatic}) are ramped across the $\mathcal{PT}$-symmetry-breaking bubble and then back to the original values. The final state subsequently features an equal population of the two eigenstates when the original Hermitian Hamiltonian has the gap closing, regardless of the initial state.

The rest part is organized as follows. In Sec. II, we present an analysis on the energy spectra of two-level systems in the presence of spin-dependent dissipations. In Sec. III, we discuss the analytical solutions of the time evolution equations. The numerical simulations of analytical results in terms of the identification of gap closing are given in Sec. IV., a brief summary is given in Sec. V.

\begin{figure}
  \centering
  \includegraphics[width=8cm]{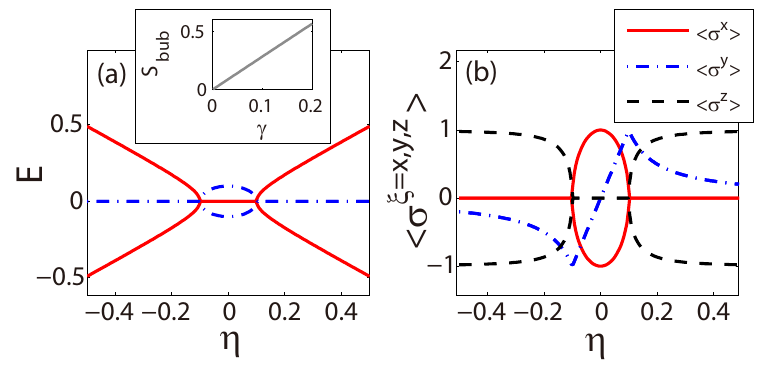}\\
  \caption{Illustration of $\mathcal{PT}$-symmetry breaking in single-spin system $H_{0}$. (a)(b): The spectra (a) and spin structure (b) of a single-spin system with $\mathcal{PT}$-symmetry breaking. The red solid (blue dash-dotted) curves in (a) denote the real (imaginary) parts of the spectra. The variation of bubble size (horizontal diameter) with $\gamma$ is shown in the inset of (a). }\label{fig:PT_few_spins}
\end{figure}

\begin{table*}
\begin{tabular}{|c|c|c|c|c|}
\hline
 & $\delta_{x}=\delta_{y}=0$ & $\delta_{x}=0$, $\delta_{y}\neq0$ & $\delta_{x}\neq0$, $\delta_{y}=0$ & $\delta_{x}\neq0$, $\delta_{y}\neq0$\tabularnewline
\hline
\hline
$\mathcal{PT}$ & Yes & Yes & No & No\tabularnewline
\hline
energy gap of $H_{0}$ varies with $\eta$ & gap closing & anticrossing & anticrossing & anticrossing\tabularnewline
\hline
\end{tabular}
\caption{The $\mathcal{PT}$ symmetry and the evolution of energy gap when tuning the gap-control parameter $\eta$ of the single-spin Hamiltonian under different parametric setting.}\label{tab:PT_gap}
\end{table*}

\section{Spectra of two-level systems with $\mathcal{PT}$-symmetric dissipations}
{We are interested in the dynamics of the two-level system $H=H_{0}+H_{p}$ with the Hermitian model $H_{0}=\eta\sigma_{z}+\delta_{x}\sigma_{x}+\delta_{y}\sigma_{y}$ and spin-dependent dissipation $H_{p}=i\gamma\sigma_{x}$, where $\eta$, $\gamma$, $\delta_{x}$ and $\delta_{y}$ are real numbers. With spin rotation around the y axis $\sigma_{x,z}\rightarrow\pm\sigma_{z,x}$, the perturbation term becomes proportional to $\sigma_{z}$ and can be implemented with state-dependent loss, which has been realized on different platforms~\cite{li2019observation, wu2019observation, ren2022chiral, naghiloo2019quantum}. The experimental implementation has been discussed in detail in Appendix B. As the dynamic behaviors of quantum systems in general are associated with their spectral structures, we will first introduce the spectral properties of these non-Hermitian two-level systems in this section.}

Symmetry $\mathcal{PT}$ is defined as the product of parity symmetry $\mathcal{P}=\sigma_{x}$ and time-reversal symmetry $\mathcal{T}=i\mathcal{K}\sigma_{y}$ with the complex conjugate operator $\mathcal{K}$, ie, $\mathcal{PT}=-\mathcal{K}\sigma_{z}$. {When $\delta_{x}=0$, $H$, $H_{0}$ and $H_{p}$ all possess the $\mathcal{PT}$ symmetry, and $H$ describes a generic $\mathcal{PT}$-symmetric model. The two-level nondissipative model shows a gap-closing transition at $\eta=0$ is characterized by $H_{0}=\eta\sigma_{z}$ , when tuning the gap control parameter $\eta$ across the zero point.} When the terms $\sigma_{x}$ and $\sigma_{y}$ are turned on, the nondissipative model shows level anticrossing, although only when $\delta_{x}=0$ the model possesses $\mathcal{PT}$ symmetry. The corresponding symmetries and spectral characteristics are summarized in Tab.~\ref{tab:PT_gap}.

As predicted by the perturbation analysis presented in Appendix A, by fixing $\gamma$ and tuning $\eta$, the transitions of $\mathcal{PT}$ symmetry in the eigenstates are observed once $H_{0}$ also possesses $\mathcal{PT}$ symmetry, reflected in the transitions between real and imaginary spectra, as exemplified in Fig.~\ref{fig:PT_few_spins}(a) with $\delta_{x}=\delta_{y}=0$. As discussed in the following, the spectral transition also emerges in another case with $\mathcal{PT}$ symmetry: $\delta_{x}=0$ and $\delta_{y}\neq 0$, but this gives rise to anticrossing rather than gap closing when the dissipative perturbation is absent.

As shown in Fig.~\ref{fig:PT_few_spins}(a), a $\mathcal{PT}$-symmetry-breaking bubble is observed around $\eta=0$ in the case where $\delta_{x}=\delta_{y}=0$ (also in the case with $\delta_{x}=0$ and $\delta_{y}\neq 0$). The expectations of Pauli matrix $\sigma^{z}$ ($\sigma^{x}$) disappear when $\gamma<|\eta|$ ($\gamma>|\eta|$), accompanying the closure of real spectra, although that of $\sigma^{y}$ is always finite throughout the phase transition [Fig.~\ref{fig:PT_few_spins}(b)]. It implies that the threshold for triggering $\mathcal{PT}$ -symmetry breaking should be determined by comparing the energy gap between two eigenstates with opposite spin polarizations of $H_{0}$ (associated with $\langle\sigma_{z}\rangle$) and the imaginary perturbation (associated with $\langle\sigma_{x}\rangle$). This point is consistent with the linear dependence of bubble size on $\gamma$, as shown in the inset of Fig.~\ref{fig:PT_few_spins}(a).

The emergence of the $\mathcal{PT}$-symmetry-breaking bubble can be formally understood with the perturbation theory discussed in Appendix A. The total Hamiltonian $H=H_{0}+H_{p}$ has the $\mathcal{PT}$ symmetry $\mathcal{PT}=\mathcal{K}\sigma^{z}$. For the Hermitian single-spin system $H_{0}$, if we set the imaginary $\sigma^{x}$ term as a perturbation when $\gamma\ll \eta$, it leads to a second-order correction $\gamma^2/(2\eta)$ [-$\gamma^2/(2\eta)$] for the lower (upper) non-perturbation eigenvalues of $H_{0}$, as we discussed in the above section. On the other hand, when $\eta\ll \gamma$, a second-order correction $i\eta^2/(2\gamma)$ [-i$\eta^2/(2\gamma)$] is added to the imaginary lower (upper) non-perturbation eigenvalues of $H_{p}$ by the real $\sigma^{z}$ term as a perturbation instead. Approaching the EP points, the energy gap monotonously decreases and finally closes at the EP points to smoothly connect the real and imaginary spectra, although the perturbation conditions gradually become invalid. We would like to note that, again, this is completely different from the conventional Hermitian systems, where a perturbation coupling between two energy levels usually opens/enlarges a gap rather than leads to gap closing.

{It is worth noting that we mainly focus on the microscopic mechanism for the emergence of EPs and imaginary-spectrum bubbles from the point of perturbation theory here (e.g., answer why the spin-dependent dissipation term leads to the decrease/increase of real/imaginary energy gap), although this phenomenon itself is already known for the community~\cite{budich2019symmetry,bergholtz2021exceptional,staalhammar2021classification,delplace2021symmetry,mandal2021symmetry}. However, previous literatures mainly focused on the symmetry protection and topological properties of nodal line ended with EPs (or EP surfaces enclosing imaginary spectra, i.e., the $\mathcal{PT}$-symmetry-breaking bubble in high dimension) on energy bands by directly diagonalizing the quasi-momentum Hamiltonians of lattice models~\cite{budich2019symmetry,bergholtz2021exceptional,staalhammar2021classification,delplace2021symmetry,mandal2021symmetry}}.

\section{Nonadiabatic transitions}
{The focus of this work is the dynamical time evolution of non-Hermitian Hamiltonians first introduced in the above section. The dynamical process is nonadiabatic, provided the gap control parameters $\eta(t)=\alpha t$ are linearly tuning through the EPs and $\mathcal{PT}$-breaking bubble without real energy gap. The observable involved is the asymptotic transition probability between the two levels for $t\rightarrow\infty$ in the slow tuning speed limit of $\eta$, that is, $\alpha\rightarrow 0$, which is also termed the adiabatic limit~\cite{longstaff2019nonadiabatic}. In the absence of a non-Hermitian term, the nonadiabatic transition is reduced to the celebrated Landau-Zener transition of two-level systems.

In previous work~\cite{longstaff2019nonadiabatic}, the authors studied the dynamical evolution of a typical $\mathcal{PT}$ symmetric model $H=\eta \sigma_{z}+i\gamma\sigma_{x}$ and predicted that the asymptotic behaviors of the transition probability in the long-time limit $t\rightarrow\infty$ show an equal distribution when $\eta$ is slowly tuned. In other words, the long-time occupation probabilities of the two levels become the same in the adiabatic limit. It has been shown numerically that the equal-distribution behavior is independent on initial state. However, a generic $\mathcal{PT}$ symmetric two-level model may include a relevant $\delta_{y}$ term, that is, $H=\eta \sigma_{z}+\delta_{y}\sigma_{y}+i\gamma\sigma_{x}$, after ignoring an unimportant diagonal term $\delta_{0}\sigma_{0}$~\cite{okugawa2019topological}. Then it is natural to ask what will happen when the $\delta_{y}$ term is turned on, which drives us to extend the theory first developed in Ref.~\cite{longstaff2019nonadiabatic}.

As shown below, we find that the asymptotic behavior of qual distribution is actually absent for the case with $\delta_{y}\neq 0$, although the symmetry $\mathcal{PT}$ is preserved and EPs also exist. Instead, the ratio between the final occupation probabilities of the two levels is given by $r_p=|(\gamma-\delta_{y})/(\gamma+\delta_{y})|$, which is independent of the initial states. It should be noted that the independence of the initial state has only been proven numerically for the case without $\delta_{y}$ term in the previous work~\cite{longstaff2019nonadiabatic}. In contrast, our results prove this point analytically for the generic $\mathcal{PT}$-symmetric model $H=\eta \sigma_{z}+\delta_{y}\sigma_{y}+i\gamma\sigma_{x}$ here. For comparison, we also briefly discuss the case without $\mathcal{PT}$ symmetry, i.e., when $\delta_{x}\neq 0$, where the initial state-independent asymptotic behavior is absent.

The route is to solve the time evolution equation directly with $\eta=\alpha t$ below. Fortunately, the time evolution equations of the generic $\mathcal{PT}$-symmetric model $H=\eta \sigma_{z}+\delta_{y}\sigma_{y}+i\gamma\sigma_{x}$ can be reduced to the celebrated parabolic cylinder equations (Weber equations)~\cite{abramowitz1988handbook}. Although the solutions are associated with the complicate confluent hypergeometric functions (Kummer functions), the asymptotic behavior we concern can be obtained with the conclusions derived by the mathematician. In the following, we will discuss the generic $\mathcal{PT}$-symmetric case $\delta_{y}\neq 0$, $\delta_{x}=0$ with $\mathcal{PT}$ symmetry first, and further discuss the case $\delta_{x}\neq 0$ and $\delta_{y}\neq 0$ without $\mathcal{PT}$ symmetry for comparison.

\subsection{Generic $\mathcal{PT}$-symmetric model: $\delta_{y}\neq 0$ and $\delta_{x}=0$}
In this case, the Hamiltonian $H=\eta(t)\sigma_{z}+\delta_{y}\sigma_y+i\gamma\sigma_{x}$ possesses the $\mathcal{PT}$ symmetry. As shown in Tab.~\ref{tab:PT_gap}, the energy gap of the Hermitian zeroth-order Hamiltonian $H_{0}=\eta(t)\sigma_{z}+\delta_{y}\sigma_y$ is closing only when $\delta_{y}=0$. In general, the level anti-crossing appears when $\delta_{y}\neq 0$ for $H_{0}$. But regardless of whether the gap of $H_{0}$ closes or not, the loss of initial state information all appears in the presence of a dissipative perturbation.

The two-level model is described by a two-component wave function $\Psi=(\psi_1,\psi_2)^T$. The time evolution equation of $\Psi$ is given by
\begin{equation}\label{eq:TE}
i\partial_{t}\left(\begin{array}{c}
\psi_{1}\\
\psi_{2}
\end{array}\right)=\left(\begin{array}{cc}
-\alpha t & i(\gamma-\delta_{y})\\
i(\gamma+\delta_{y}) & \alpha t
\end{array}\right)\left(\begin{array}{c}
\psi_{1}\\
\psi_{2}
\end{array}\right).
\end{equation}
This is a group of coupled first-order differential equations. Fortunately, this equation group can be decoupled by taking the second derivative on the two side. Specifically, the decoupled second-order differential equation is given by $i\partial_{t}^2\Psi=(\partial_{t}H-iH^2)\Psi$, that is,
\begin{equation}\label{eq:2order_H}
i\partial_{t}^{2}\psi_{1,2}=\{\mp\alpha-i[(\alpha t)^{2}-(\gamma^{2}-\delta_{x}^{2}-\delta_{y}^{2})+2i\delta_{x}\gamma]\}\psi_{1,2}.
\end{equation}
These equations are actually the standard forms of the parabolic cylinder equations (Weber equations)~\cite{abramowitz1988handbook}
\begin{equation}\label{eq:PCE}
\frac{d^{2}\psi_{\nu}^{2}}{dz^{2}}-(\frac{1}{4}z^{2}+a_{\nu})\psi_{\nu}=0,\quad\nu=1,2,
\end{equation}
by defining the new argument $z=e^{-i\pi/4}\sqrt{2\alpha}t$ and constants
\begin{equation}\label{eq:a_12}
a_{1,2}=\frac{i(\gamma^{2}-\delta_{y}^{2})}{2\alpha}\mp\frac{1}{2},
\end{equation}
and can be solved analytically with the confluent hypergeometric functions (i.e., the Kummer's functions). In the following, we will first discuss the special solutions of Eq.~(\ref{eq:a_12}). Then a generic solution can be written with the combination of these special solutions. The combination coefficients can be fixed with the initial conditions (i.e., the initial states and the initial first-order derivative of wave functions). Finally, the asymptotic behaviors of the wave function can be derived with the asymptotic behaviors of the confluent hypergeometric functions. Although the procedure is tedious, the idea is very straightforward.

Before moving to the solutions of the evolution equation, let us first give some comments on the difference with the analysis on the special $\mathcal{PT}$-symmetric model in Ref.~\cite{longstaff2019nonadiabatic}. In fact, the form of time evolution equations are all the parabolic cylinder equations shown in Eq.~(\ref{eq:PCE}). Only the forms of $a_{j=1,2}$ are different, as shown in Eq.~(\ref{eq:a_12}). According to the theory in the previous work~\cite{longstaff2019nonadiabatic}, it actually implies the nonadiabatic transitions also show equal-distribution asymptotic behavior even for the case with $\delta_{y}\neq 0$ discussed here. However, we will show below that, the equal-distribution asymptotic behavior actually is absent in the current case, which is also confirmed in our numerics.
}

The parabolic cylinder equations Eq.~(\ref{eq:PCE}) all have two exact symmetric particular solutions, i.e., the even-parity solution,
\begin{equation}\label{eq:y1}
\begin{split}
&y_{\nu1}=e^{-\frac{1}{4}z^{2}}M(\frac{1}{2}a_{\nu}+\frac{1}{4},\frac{1}{2},\frac{1}{2}z^{2})\\
&=e^{-\frac{1}{4}z^{2}}\{1+(a_{\nu}+\frac{1}{2})\frac{z^{2}}{2!}+(a_{\nu}+\frac{1}{2})(a_{\nu}+\frac{5}{2})\frac{z^{2}}{2!}+\cdots\},
\end{split}
\end{equation}
and the odd-parity solution,
\begin{equation}\label{eq:y2}
\begin{split}
&y_{\nu2}=ze^{-\frac{1}{4}z^{2}}M(\frac{1}{2}a_{\nu}+\frac{3}{4},\frac{3}{2},\frac{1}{2}z^{2})\\
&=e^{-\frac{1}{4}z^{2}}\{z+(a_{\nu}+\frac{3}{2})\frac{z^{3}}{3!}+(a_{\nu}+\frac{3}{2})(a_{\nu}+\frac{7}{2})\frac{z^{5}}{5!}+\cdots\},
\end{split}
\end{equation}
where $M$ functions are the confluent hypergeometric functions. Any general solutions of Eq.~(\ref{eq:TE}) are the superpositions of the two particular solutions, i.e.,
\begin{equation}\label{eq:G_solutions}
\left(\begin{array}{c}
\psi_{1}\\
\psi_{2}
\end{array}\right)=\left(\begin{array}{c}
\alpha_{11}y_{11}+\alpha_{12}y_{12}\\
\alpha_{21}y_{21}+\alpha_{22}y_{22}
\end{array}\right).
\end{equation}
{In the next step, we will solve these coefficients with the initial conditions.}

The systems are assumed to be prepared on an arbitrary initial state at $t=0$,
\begin{equation}\label{eq:initial_state}
\Psi(0)=\left(\begin{array}{c}
\psi_{1}(0)\\
\psi_{2}(0)
\end{array}\right)=\left(\begin{array}{c}
A\\
B
\end{array}\right).
\end{equation}
Considering $y_{1}(0)=1$ and $y_{2}(0)=0$, we have $\alpha_{11}=A$ and $\alpha_{21}=B$. The initial conditions of first-order differential equations give rise to
\begin{equation}\label{eq:first_order_IC}
\frac{d}{dz}\Psi(z)\bigg|_{z=0}=\frac{e^{i\pi/4}}{\sqrt{2\alpha}}\left(\begin{array}{cc}
0 & \gamma-\delta_{y}\\
\gamma+\delta_{y} & 0
\end{array}\right)\left(\begin{array}{c}
A\\
B
\end{array}\right).
\end{equation}
It is easy to confirm that $dy_{1}(0)/dz=0$ and $dy_{2}(0)/dz=1$ from the expansion series in Eq.~(\ref{eq:y1}) and Eq.~(\ref{eq:y2}). With these conditions, we have
\begin{equation}\label{eq:coefs}
\left(\begin{array}{c}
\alpha_{12}\\
\alpha_{22}
\end{array}\right)=\left(\begin{array}{c}
\frac{e^{i\pi/4}}{\sqrt{2\alpha}}(\gamma-\delta_{y})B\\
\frac{e^{i\pi/4}}{\sqrt{2\alpha}}(\gamma+\delta_{y})A
\end{array}\right).
\end{equation}
Therefore, the solutions of the time evolution equations~(\ref{eq:TE}) are given by
\begin{equation}\label{eq:TE_solution}
\Psi=\left(\begin{array}{c}
Ay_{11}+\frac{e^{i\pi/4}}{\sqrt{2\alpha}}(\gamma-\delta_{y})By_{12}\\
By_{21}+\frac{e^{i\pi/4}}{\sqrt{2\alpha}}(\gamma+\delta_{y})Ay_{22}
\end{array}\right).
\end{equation}

{We mainly concern the asymptotic behavior in the large $t$ limit. Then our target is to employ the asymptotic behaviors of $M$ functions to analyze the asymptotic behaviors of $\Psi$. Fortunately, the asymptotic behaviors of the confluent hypergeometric functions~\cite{abramowitz1988handbook} can be written down as
\begin{equation}\label{eq:M_symp}
\begin{split}
&\frac{M(a,b,x)}{\Gamma(b)}=\frac{e^{\pm i\pi a}x^{-a}}{\Gamma(b-a)}\{\sum_{n=0}^{R-1}\frac{(a)_{n}(1+a-b)_{n}}{n!}(-x)^{-n}\\
&+\mathcal{O}(|x|^{-R})\}+\frac{e^{x}x^{a-b}}{\Gamma(a)}\{\sum_{n=0}^{S-1}\frac{(b-a)_{n}(1-a)_{n}}{n!}x{}^{-n}\\
&\mathcal{O}(|x|^{-S})\},
\end{split}
\end{equation}
when $|x|\rightarrow\infty$, where
\begin{equation}\label{eq:a_n}
(f)_{n}=f(f+1)(f+2)\ldots(f+n-1),(f)_{0}=1.
\end{equation}
The signs $\pm$ correspond to the argument of $x$ in the ranges $-\frac{1}{2}\pi<\arg(x)<\frac{3}{2}\pi$ and $-\frac{3}{2}\pi<\arg(x)\leq-\frac{1}{2}\pi$, respectively.

Now let us discuss the leading order terms for the asymptotic behaviors of the generic solutions in Eqs.~(\ref{eq:G_solutions}). From Eqs.~(\ref{eq:y1}) and (\ref{eq:y2}), we can find $y_{1}$ and $y_{2}$ involves different coefficients $a$ and $b$, but the same argument $x=z^2/2$ appearing in Eq.~(\ref{eq:M_symp}). Since $\arg(z^2/2)=-\pi/2$, we always need take $-$ sign in Eq.~(\ref{eq:M_symp}). The leading order terms of $y$ functions are determined by the magnitudes of $-a$ and $(a-b)$, since only the zero-order terms of the expansion series in Eq.~(\ref{eq:M_symp}) needs to be considered. The values of $-a$ and $(a-b)$ are listed in Tab.~\ref{tab:a_b}. It's shown that, for $y_{11}$ and $y_{12}$ ($y_{21}$ and $y_{22}$), the first terms with power $-a$ (the second terms with power $(a-b)$) in Eq.~(\ref{eq:M_symp}) are dominant. Relative to $y_{11}$ and $y_{21}$, although $y_{12}$ and $y_{22}$ have additional $x$ argument before the $M$ function from Eqs~(\ref{eq:y1}) and (\ref{eq:y2}), they are counteracted by the real part of $-a$ and $(a-b)$. Therefore, the asymptotic behaviors of $y$ functions are all proportional zero-order term of $z$, which are given by
\begin{widetext}
\begin{equation}\label{eq:ab_y}
\begin{split}
&y_{11}\thicksim e^{-\frac{1}{4}x^{2}}\frac{\Gamma(\frac{1}{2})e^{\pi(\gamma^{2}-\delta_{y}^{2})/4\alpha}}{\Gamma(\frac{1}{2}-\frac{i(\gamma^{2}-\delta_{y}^{2})}{4\alpha})}(\frac{x^{2}}{2})^{-i(\gamma^{2}-\delta_{y}^{2})/4\alpha},\quad y_{12}\thicksim-ie^{-\frac{1}{4}x^{2}}\frac{\sqrt{2}\Gamma(\frac{3}{2})e^{\pi(\gamma^{2}-\delta_{y}^{2})/4\alpha}}{\Gamma(1-\frac{i(\gamma^{2}-\delta_{y}^{2})}{4\alpha})}(\frac{x^{2}}{2})^{-i(\gamma^{2}-\delta_{y}^{2})/4\alpha},\\
&y_{21}\thicksim e^{\frac{1}{4}x^{2}}\frac{\Gamma(\frac{1}{2})}{\Gamma(\frac{1}{2}+\frac{i(\gamma^{2}-\delta_{y}^{2})}{4\alpha})}(\frac{x^{2}}{2})^{i(\gamma^{2}-\delta_{y}^{2})/4\alpha},\quad y_{22}\thicksim e^{\frac{1}{4}x^{2}}\frac{\sqrt{2}\Gamma(\frac{3}{2})}{\Gamma(1+\frac{i(\gamma^{2}-\delta_{y}^{2})}{4\alpha})}(\frac{x^{2}}{2})^{i(\gamma^{2}-\delta_{y}^{2})/4\alpha}.
\end{split}
\end{equation}

Further, by substituting the expressions in Eq~(\ref{eq:ab_y}) into Eq.~(\ref{eq:TE_solution}), we yield
\begin{equation}\label{eq:ratio}
\begin{split}
\frac{\psi_{1}}{\psi_{2}}\sim & e^{\pi(\gamma^{2}-\delta_{y}^{2})/4\alpha}(\frac{z^{2}}{2})^{-\frac{i(\gamma^{2}-\delta_{y}^{2})}{2\alpha}}[A\frac{\Gamma(\frac{1}{2})}{\Gamma(\frac{1}{2}-\frac{i(\gamma^{2}-\delta_{y}^{2})}{4\alpha})}+Be^{-i\pi/4}\frac{\gamma-\delta_{y}}{\sqrt{\alpha}}\frac{\Gamma(\frac{3}{2})}{\Gamma(1-\frac{i(\gamma^{2}-\delta_{y}^{2})}{4\alpha})}]\\
&/[B\frac{\Gamma(\frac{1}{2})}{\Gamma(\frac{1}{2}+\frac{i(\gamma^{2}-\delta_{y}^{2})}{4\alpha})}+Ae^{i\pi/4}\frac{\gamma+\delta_{y}}{\sqrt{\alpha}}\frac{\Gamma(\frac{3}{2})}{\Gamma(1+\frac{i(\gamma^{2}-\delta_{y}^{2})}{4\alpha})}].
\end{split}
\end{equation}
\end{widetext}
By defining $\tilde{\gamma}^2=(\gamma^2-\delta_y^2)/\alpha$ and considering $\Gamma(1/2)=2\Gamma(3/2)$, we have
\begin{equation}\label{eq:ratio_abs}
\begin{split}
&\left|\frac{\psi_{1}}{\psi_{2}}\right|\sim\frac{e^{\pi\frac{\tilde{\gamma}^{2}}{4}}\left|(\frac{z^{2}}{2})^{-\frac{i\tilde{\gamma}^{2}}{2}}\right|\cdot\left|\frac{2A}{\Gamma(\frac{1}{2}-\frac{i\tilde{\gamma}^{2}}{4})}+\frac{\gamma-\delta_{y}}{\sqrt{\alpha}}\frac{e^{-i\pi/4}B}{\Gamma(1-\frac{i\tilde{\gamma}^{2}}{4})}\right|}{\left|\frac{2B}{\Gamma(\frac{1}{2}+\frac{i\tilde{\gamma}^{2}}{4})}+\frac{\gamma+\delta_{y}}{\sqrt{\alpha}}\frac{e^{i\pi/4}A}{\Gamma(1+\frac{i\tilde{\gamma}^{2}}{4})}\right|}\\
&=\left|\frac{2\sqrt{\alpha}\Gamma(1+\frac{i\tilde{\gamma}^{2}}{4})e^{\pi\frac{\tilde{\gamma}^{2}}{4}}(\frac{z^{2}}{2})^{-\frac{i\tilde{\gamma}^{2}}{2}}}{(\gamma+\delta_{y})\Gamma(\frac{1}{2}-\frac{i\tilde{\gamma}^{2}}{4})}\right|\\
&\times \left|\frac{A+Be^{-i\pi/4}\frac{\gamma-\delta_{y}}{2\sqrt{\alpha}}\frac{\Gamma(\frac{1}{2}-\frac{i\tilde{\gamma}^{2}}{4})}{\Gamma(1-\frac{i\tilde{\gamma}^{2}}{4})}}{A+Be^{-i\pi/4}\frac{2\sqrt{\alpha}}{\gamma+\delta_{y}}\frac{\Gamma(1+\frac{i\tilde{\gamma}^{2}}{4})}{\Gamma(\frac{1}{2}+\frac{i\tilde{\gamma}^{2}}{4})}}\right|.
\end{split}
\end{equation}
Since $\Gamma(\xi)=\int_{0}^{\infty}d\tau\tau^{\xi-1}e^{-\tau}$, $\Gamma(\xi)^\ast=\Gamma(\xi^\ast)$, and then the ratio between the coefficients before $B$ in the above equation is given by
\begin{equation}\label{eq:ratio_B}
\frac{\tilde{\gamma}^{2}}{4}\frac{\Gamma(\frac{1}{2}-\frac{i\tilde{\gamma}^{2}}{4})\Gamma(\frac{1}{2}+\frac{i\tilde{\gamma}^{2}}{4})}{\Gamma(1-\frac{i\tilde{\gamma}^{2}}{4})\Gamma(1+\frac{i\tilde{\gamma}^{2}}{4})}=\frac{\tilde{\gamma}^{2}}{4}\frac{\left|\Gamma(\frac{1}{2}+\frac{i\tilde{\gamma}^{2}}{4})\right|^{2}}{\left|\Gamma(1+\frac{i\tilde{\gamma}^{2}}{4})\right|^{2}}.
\end{equation}
Further, by employing the properties of Gamma function,
\begin{equation}\label{eq:GF_beh}
\left|\Gamma(\frac{1}{2}+\lambda i)\right|^{2}=\frac{\pi}{\cosh(\pi\lambda)},\quad\left|\Gamma(1+\lambda i)\right|^{2}=\frac{\pi\lambda}{\sinh(\pi\lambda)},
\end{equation}
the above equation can be rewritten as
\begin{equation}\label{eq:ratio_B_1}
\frac{\tilde{\gamma}^{2}}{4}\frac{\left|\Gamma(\frac{1}{2}+\frac{i\tilde{\gamma}^{2}}{4})\right|^{2}}{\left|\Gamma(1+\frac{i\tilde{\gamma}^{2}}{4})\right|^{2}}=\tanh(\pi\frac{\tilde{\gamma}^{2}}{4}).
\end{equation}
In the adiabatic limit $\tilde{\gamma}^2\rightarrow\infty$, this ratio becomes $1$. Thus, we finally have
\begin{equation}\label{eq:ratio_psi}
\begin{split}
&\lim_{\tilde{\gamma}\rightarrow\infty}\left|\frac{\psi_{1}}{\psi_{2}}\right|\sim\lim_{\tilde{\gamma}\rightarrow\infty}\left|\frac{2\sqrt{\alpha}\Gamma(1+\frac{i\tilde{\gamma}^{2}}{4})e^{\pi\frac{\tilde{\gamma}^{2}}{4}}(\frac{z^{2}}{2})^{-\frac{i\tilde{\gamma}^{2}}{2}}}{(\gamma+\delta_{y})\Gamma(\frac{1}{2}-\frac{i\tilde{\gamma}^{2}}{4})}\right|\\
&=\lim_{\tilde{\gamma}\rightarrow\infty}\left|\frac{2\sqrt{\alpha}}{\gamma+\delta_{y}}e^{\pi\frac{\tilde{\gamma}^{2}}{4}}(e^{-i\frac{\pi}{2}}\frac{\left|z\right|^{2}}{2})^{-\frac{i\tilde{\gamma}^{2}}{2}}\right|\cdot\left|\frac{\tilde{\gamma}^{2}}{4\tanh(\pi\tilde{\gamma}^{2}/4)}\right|^{1/2}\\
&=\left|\frac{\gamma-\delta_{y}}{\gamma+\delta_{y}}\right|^{1/2}.
\end{split}
\end{equation}
It proves that the ratio between the occupation probabilities of the two states is a constant $r_p=|(\gamma-\delta_{y})/(\gamma+\delta_{y})|$, which is independent of the initial states. Besides, only when $\delta_{y}=0$, $r_p=1$, which proves the equal-distribution asymptotic behavior elaborated at the beginning of this section. Although the $\mathcal{PT}$-symmetry-breaking bubble also emerges for finite $\delta_{y}$, the equal-distribution asymptotic behavior does not exist in this case.
}

\begin{table}
\centering
\begin{tabular}{|c|c|c|c|c|}
\hline
 & $a$ & $b$ & $-a$ & $a-b$\tabularnewline
\hline
\hline
$y_{11}$ & $\frac{1}{2}a_{1}+\frac{1}{4}$ & $\frac{1}{2}$ & $-\frac{i(\gamma^{2}-\delta_{y}^{2})}{4\alpha}$ & $\frac{i(\gamma^{2}-\delta_{y}^{2})}{4\alpha}-\frac{1}{2}$\tabularnewline
\hline
$y_{12}$ & $\frac{1}{2}a_{1}+\frac{3}{4}$ & $\frac{3}{2}$ & $-\frac{i(\gamma^{2}-\delta_{y}^{2})}{4\alpha}-\frac{1}{2}$ & $\frac{i(\gamma^{2}-\delta_{y}^{2})}{4\alpha}-1$\tabularnewline
\hline
$y_{21}$ & $\frac{1}{2}a_{2}+\frac{1}{4}$ & $\frac{1}{2}$ & $-\frac{i(\gamma^{2}-\delta_{y}^{2})}{4\alpha}-\frac{1}{2}$ & $\frac{i(\gamma^{2}-\delta_{y}^{2})}{4\alpha}$\tabularnewline
\hline
$y_{22}$ & $\frac{1}{2}a_{2}+\frac{3}{4}$ & $\frac{3}{2}$ & $-\frac{i(\gamma^{2}-\delta_{y}^{2})}{4\alpha}-1$ & $\frac{i(\gamma^{2}-\delta_{y}^{2})}{4\alpha}-\frac{1}{2}$\tabularnewline
\hline
\end{tabular}
\caption{The values of $-a$ and $(a-b)$ for different $y$ functions.}\label{tab:a_b}
\end{table}

\subsection{Case without $\mathcal{PT}$ symmetry: $\delta_{y}\neq 0$ and $\delta_{x}\neq 0$}
When $\delta_{x}\neq 0$, no matter $\delta_{y}$ is finite or not, the $\mathcal{PT}$ symmetry is absent. We study this case just for comparison. In this case, the time evolution equation is given by
\begin{equation}\label{eq:TE_sgmx}
i\partial_{t}\left(\begin{array}{c}
\psi_{1}\\
\psi_{2}
\end{array}\right)=\left(\begin{array}{cc}
-\alpha t & \delta_{x}+i(\gamma-\delta_{y})\\
\delta_{x}+i(\gamma+\delta_{y}) & \alpha t
\end{array}\right)\left(\begin{array}{c}
\psi_{1}\\
\psi_{2}
\end{array}\right).
\end{equation}
The second-order differential equation takes the form
\begin{equation}\label{eq:second_sgmx}
i\partial_{t}^{2}\psi_{1,2}=\{\mp\alpha-i[(\alpha t)^{2}-(\gamma^{2}-\delta_{x}^{2}-\delta_{y}^{2})+2i\delta_{x}\gamma]\}\psi_{1,2},
\end{equation}
which corresponds to the parabolic cylinder equation
\begin{equation}\label{eq:web_sgmx}
\frac{d^{2}\psi_{\xi}^{2}}{dz^{2}}-(\frac{1}{4}z^{2}+a_{\xi})\psi_{\xi}=0,\quad\xi=1,2,
\end{equation}
with
\begin{equation}\label{eq:as_sgmx}
a_{1,2}=\frac{i(\gamma^{2}-\delta_{x}^{2}-\delta_{y}^{2})+2\delta_{x}\gamma}{2\alpha}\mp\frac{1}{2},
\end{equation}
as well. The only difference is the form of $a$ factor.

For simplicity, we take a special initial state $\Psi(0)=(\begin{array}{cc}
0 & 1\end{array})^{T}$ as an example. Similar to the above subsection, with this initial conditions, it gives rise to the solution
\begin{equation}\label{eq:solution_sgmx}
\begin{split}
&\Psi=\left(\begin{array}{c}
e^{-i\pi/4}\frac{\delta_{x}+i(\gamma-\delta_{y})}{\sqrt{2\alpha}}y_{12}\\
y_{21}
\end{array}\right)\\
&=e^{-\frac{1}{4}z^{2}}\left(\begin{array}{c}
e^{-i\pi/4}\frac{\delta_{x}+i(\gamma-\delta_{y})}{\sqrt{2\alpha}}zM(\frac{1}{2}a_{1}+\frac{3}{4},\frac{3}{2},\frac{1}{2}z^{2})\\
M(\frac{1}{2}a_{2}+\frac{1}{4},\frac{1}{2},\frac{1}{2}z^{2})
\end{array}\right).
\end{split}
\end{equation}

In the limit $z\rightarrow\infty$, for $\psi_{1}$, because
\begin{equation}\label{eq:a_b_psi1}
\begin{split}
&-a=-(\frac{a_{1}}{2}+\frac{3}{4})=-\frac{1}{2}-\frac{i(\gamma^{2}-\delta_{y}^{2}-\delta_{x}^{2})+2\delta_{x}\delta_{y}}{4\alpha}\\
&a-b=\frac{a_{1}}{2}-\frac{3}{4}=-1+\frac{i(\gamma^{2}-\delta_{y}^{2}-\delta_{x}^{2})+2\delta_{x}\delta_{y}}{4\alpha},
\end{split}
\end{equation}
all have large real parts $\delta_{x}\delta_{y}/\alpha$ in the adiabatic limit, we need to compare their real parts in order to determine the dominant terms in the asymptotic expansion of $M$ functions [see Eq.~(\ref{eq:M_symp})]. It can be divided into two cases, when $\delta_{y}\delta_{x}/\alpha<0.5$, the first term in Eq.~(\ref{eq:M_symp}) is dominant; when $\delta_{y}\delta_{x}/\alpha>0.5$, the second term becomes dominant. For $\psi_{2}$, similarly, we have
\begin{equation}\label{eq:a_b_psi2}
\begin{split}
&-a=-(\frac{a_{2}}{2}+\frac{1}{4})=-\frac{1}{2}-\frac{i(\gamma^{2}-\delta_{y}^{2}-\delta_{x}^{2})+2\delta_{x}\delta_{y}}{4\alpha}\\
&a-b=\frac{a_{2}}{2}-\frac{1}{4}=\frac{i(\gamma^{2}-\delta_{y}^{2}-\delta_{x}^{2})+2\delta_{x}\delta_{y}}{4\alpha}.
\end{split}
\end{equation}
When $\delta_{y}\delta_{x}/\alpha<-0.5$, the first term in the asymptotic expansion is dominant, but when $\delta_{y}\delta_{x}/\alpha>-0.5$, the second term becomes dominant. These results are reasonable, because the EPs are absent when $\delta_{x}\neq 0$ and the two levels are always gapped when tuning $\eta$, and thus the time evolution becomes adiabatic in the slow-tuning-speed limit.

In summary, when $\delta_{y}\delta_{x}/\alpha<-0.5$, by defining
\begin{equation}\label{eq:epsilon}
\epsilon=\frac{i(\gamma^{2}-\delta_{y}^{2}-\delta_{x}^{2})+2\delta_{x}\delta_{y}}{4\alpha},
\end{equation}
we have
\begin{equation}\label{eq:case1}
\begin{split}
&M(\frac{1}{2}a_{1}+\frac{3}{4},\frac{3}{2},\frac{1}{2}z^{2})=\frac{\Gamma(\frac{3}{2})e^{-i\pi(\frac{1}{2}+\epsilon)}(\frac{z^{2}}{2})^{-(\frac{1}{2}+\epsilon)}}{\Gamma(1-\epsilon)}\\
&M(\frac{1}{2}a_{2}+\frac{1}{4},\frac{1}{2},\frac{1}{2}z^{2})=\frac{\Gamma(\frac{1}{2})e^{-i\pi(\frac{1}{2}+\epsilon)}(\frac{z^{2}}{2})^{-(\frac{1}{2}+\epsilon)}}{\Gamma(-\epsilon)},
\end{split}
\end{equation}
which implies $\psi_1\propto z^{-2\epsilon}$ and $\psi_1\propto z^{-1-2\epsilon}$ and thus $\left|\frac{\psi_{1}}{\psi_{2}}\right|^{2}\propto z^{2}\rightarrow\infty$.
Similarly, when $-0.5<\delta_{y}\delta_{x}/\alpha<0.5$, $\psi_1\propto z^{-2\epsilon}$ and $\psi_1\propto z^{-2\epsilon}$, which implies only when $\delta_{y}\delta_{z}=0$, it is possible to get $\left|\frac{\psi_{1}}{\psi_{2}}\right|^{2}\rightarrow1$. As we shown in the above subsection, actually, only when $\delta_{y}=\delta_{z}=0$, we have this result, i.e., the equal-distribution asymptotic behavior.
For the case $\delta_{y}\delta_{x}/\alpha>0.5$, we have $\psi_1\propto z^{-2-2\epsilon}$ and $\psi_1\propto z^{2\epsilon}$, $\left|\frac{\psi_{1}}{\psi_{2}}\right|^{2}$ also can not give rise to the equal-distribution asymptotic behavior.

\begin{figure*}
  \centering
  \includegraphics[width=16cm]{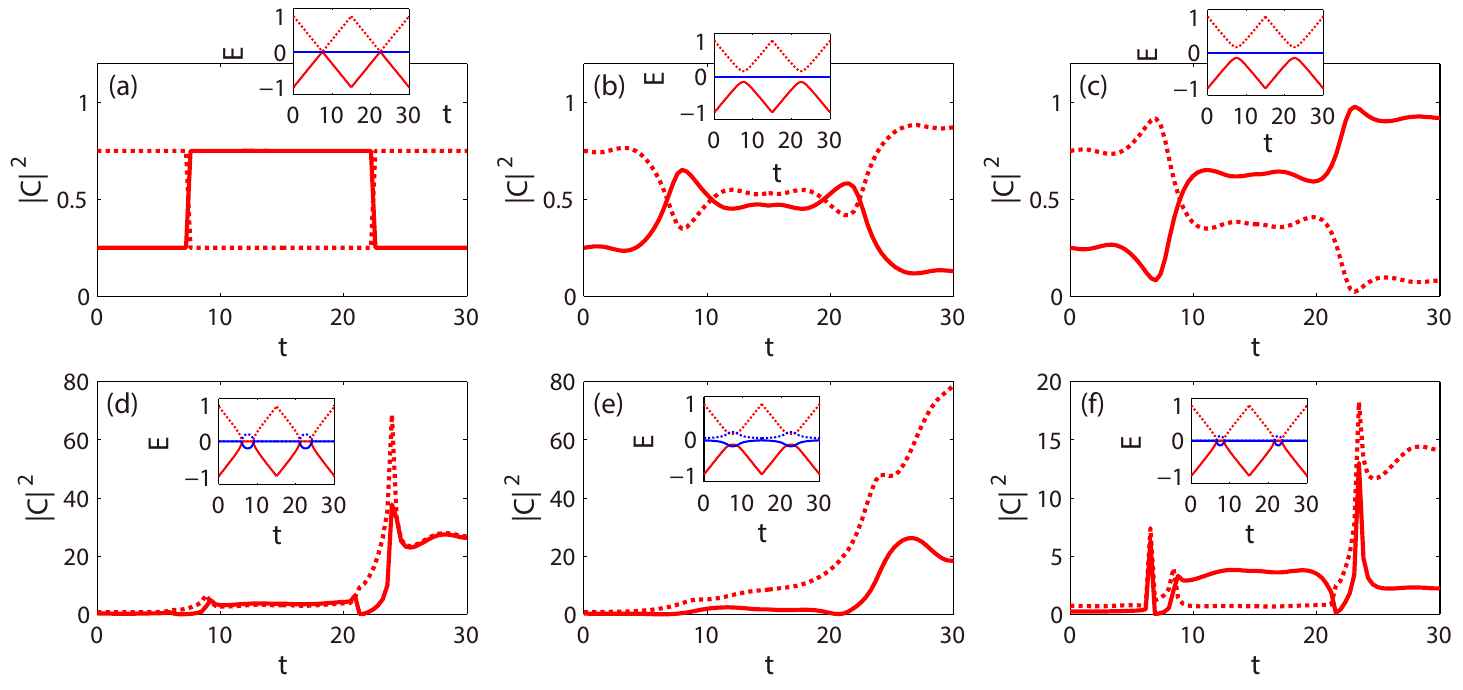}\\
  \caption{The illustration of instantaneous spectra (upper row) and projection properties (lower row) for different perturbation parameters in a cyclic time evolution, where $\eta(t)=-1+\alpha t$ when $t<t_{f}$ and $\eta(t)=1-\alpha t$ when $t>t_{f}$ with the one-way evolution time $t_{f}=15$ and tuning speed $\alpha=0.025$. The initial states are taken as $|\Psi(0)\rangle=\cos(\theta)|\Psi_{1}(0)\rangle+e^{i\varphi}\sin(\theta)|\Psi_{2}(0)\rangle$ with $\theta=\pi/3$ and $\varphi=\pi/6$, where $|\Psi_{1,2}(0)\rangle$ are the instantaneous eigenstates at moment $t=0$. The spectra of real-time Hamiltonian $H=\eta(t)\sigma_{z}+\delta_{x}\sigma_x+\delta_{y}\sigma_y+i\gamma\sigma_x$ (insets) and the projection probabilities of instantaneous states when $\gamma=\delta_{x}=\delta_{y}=0$ (a) [$\delta_{y}=\delta_{x}=0$, $\gamma=0.2$ (d)], $\gamma=\delta_{y}=0$, $\delta_{x}=0.15$ (b) [$\delta_{y}=0$, $\delta_{x}=0.15$, $\gamma=0.2$ (e)], and $\gamma=\delta_{x}=0$, $\delta_{y}=0.15$ (c) [$\delta_{x}=0$, $\delta_{y}=0.15$, $\gamma=0.2$ (e)], are shown. The red (blue) curves in the insets represent the real (imaginary) parts of the spectra.  The solid (dotted) curves in the sub-figures show the projection probabilities for the corresponding levels in the insets. The projection probabilities $|C_{\xi=1,2}|^2$ are defined as $C_{1,2}=\langle\tilde{\Psi}_{1,2}(t)|\Psi(t)\rangle/|\langle\tilde{\Psi}_{1,2}|\Psi_{1,2}\rangle|$ with the instantaneous state $|\Psi(t)\rangle$ and the instantaneous right (left) eigenstates $|\Psi_{1,2}(t)\rangle$ ( $|\tilde{\Psi}_{1,2}(t)\rangle$)~\cite{brody2013biorthogonal}. We would like to note that the coefficients $C_{1,2}$ are not the spin components directly. However, because at $t=t_{f}$ $\eta$ is far larger than other coefficients in the Hamiltonian, $\Psi_{1,2}$ approach the eigenstate of $\sigma_{z}$. Therefore, the asymptotic behaviors of $C_{1,2}$ approach those of $\psi_{1,2}$ given in the above section. The introduce of $C_{1,2}$ is just for the convenience of experimental observation.}\label{fig:circle_evolution}
\end{figure*}

Therefore, although when $\delta_{y}\neq 0$, the $\mathcal{PT}$ symmetry and the initial-state-independent asymptotic behavior are all preserved, the equal-distribution asymptotic behavior disappears. In contrast, only when $\delta_{x}=\delta_{y}=0$, the nonadiabatic transitions show the equal-distribution asymptotic behavior. When $\delta_{x}\neq0$, we have adiabatic transitions because the EPs are absent. These are the main conclusions of this article.

Since for $H_{0}$, only when $\delta_{x}=\delta_{y}=0$, the variation of $\eta$ gives rise to gap closing transition. For $\delta_{x}$ or $\delta_{y}$ becomes finite, the gap closing is replaced by an anti-crossing of the two levels. Therefore, the equal-distribution asymptotic behavior may be employed as an identification of gap closing. We will show the numerical simulations in terms of this application below.

\section{Identification of gap closing with dynamical method}
As shown above, we find the equal-distribution behaviour is present when the non-dissipative Hamiltonian has a gap-closing transition, and disappears when the non-dissipative Hamiltonian displays level anti-crossing even when the $\mathcal{PT}$ symmetry and thus EPs are still present. By employing these unique properties, we propose to identify the energy gap closing in the non-dissipative Hamiltonian with a cyclic time evolution covering the the $\mathcal{PT}$-symmetry-breaking bubble, as a mimic of Landau-Zener-Stückelberg interference~\cite{shen2019landau}. It can be implemented by driving particles on the Bloch bands through a static force~\cite{longstaff2019nonadiabatic,shen2019landau}. Specifically, we propose to start from a parametric point, e.g., $\eta=-1$, and tune $\eta$ across the the $\mathcal{PT}$-symmetry-breaking bubble at around $\eta=0$ and back to the start point $\eta=-1$. By observing the projection probabilities of final state for the instantaneous ground and excited states, which are connected with the states that coalesces at the EPs, we can gain the signature of gap closing. So long as the tuning speed is slow enough, we can observe almost equal projection probabilities when $H_{0}$ indeed experiences a gap-closing transition.

The instantaneous projection probabilities and spectra (insets) of a cyclic time evolution ($\eta(t)=-1+\alpha t$ when $t<t_{f}$ and $\eta(t)=1-\alpha t$ when $t>t_{f}$ with $\alpha=1/15$) are shown in Fig.~\ref{fig:circle_evolution} for different parameters (all with arbitrary units in this paper). The upper and lower rows of subfigures show the cases without and with dissipative perturbations. The three sub-figures from left to right correspond to the Hamiltonians $H_{0}=\eta(t)\sigma_{z}$, $\eta(t)\sigma_{z}+\delta_{x}\sigma_x$ and $\eta(t)\sigma_{z}+\delta_{y}\sigma_y$ (i.e., $H=\eta(t)\sigma_{z}+i\gamma\sigma_{x}$, $\eta(t)\sigma_{z}+\delta_{x}\sigma_x+i\gamma\sigma_{x}$ and $\eta(t)\sigma_{z}+\delta_{y}\sigma_y+i\gamma\sigma_{x}$), respectively. The initial state is prepared as an arbitrary superposition of the eigenstates of the lower and higher levels at $t=0$. As shown in Fig.~\ref{fig:circle_evolution}(a), where $H=H_{0}=\eta(t)\sigma_{z}$, the instantaneous state has invariant projection probabilities for the eigenstates, although the spectra show a gap closing. It is because there is no any coupling between the ground and excited states. When the imaginary perturbation is present, i.e., $H=H_{0}+H_{p}=\eta(t)\sigma_{z}+i\gamma H_{p}$ , where a the $\mathcal{PT}$-symmetry-breaking bubble emerges at around the gap-closing point [see the inset of Fig.~\ref{fig:circle_evolution}(d)], the instantaneous projection probabilities for the two levels take almost the same instantaneous values after passing the EPs [see Fig.~\ref{fig:circle_evolution}(d)]. This is a good signature for probing the gap-closing transition in the original Hamiltonian $H_{0}$.

Let us further discuss the cases where the gap closing in $H_{0}$ is absent. Upon (a) and (d), a real spin term $\delta_{x}\sigma^{x}$ is introduced to (b) and (e). This real spin term prevents the gap closing at $\eta=0$ and leads to anticrossing at around $\eta=0$. Besides, it also breaks the $\mathcal{PT}$ symmetry, because $\mathcal{K}\sigma^{z}\sigma^{x}\sigma^{z}\mathcal{K}^{\dagger}=-\sigma^{x}$ as discussed above. In Fig.~\ref{fig:circle_evolution} (b), where the imaginary perturbation is not turned on, it is shown the spectra indeed has a finite gap when $\eta=0$ [see the inset]. The finite real energy gap allows the Landau-Zener tunneling from lower level to the higher level and the instantaneous projection probabilities unequally distributes. When the time evolution tends to the adiabatic limit $\delta_{x}^2/\alpha\gg1$ (here $\delta_{x}^2/\alpha\sim 0.34$), the projection probabilities should keep constant as the energy gap completely suppresses the tunneling. Due to the absence of $\mathcal{PT}$ symmetry, the spectra in the case with imaginary perturbation [see inset of (e)] always have imaginary parts, and the instantaneous projection probability of the level with positive imaginary spectra is amplified after long-time evolution and becomes dominant.

In Fig.~\ref{fig:circle_evolution} (c) and (f), another kind of real term $\delta_{y}\sigma^{y}$ is introduced into $H_{0}$ to break the gap closing. This perturbation also opens a finite gap at $\eta=0$ and thus leads to anticrossing as well. But unlike (b) and (e), the $\mathcal{PT}$ symmetry is still preserved in this case, since $\mathcal{K}\sigma^{z}\sigma^{y}\sigma^{z}\mathcal{K}^{\dagger}=\sigma^{y}$. The behaviours of the time evolution in the case without imaginary perturbation look like the case with $\delta_{x}$ term [see (c)]. When the time evolution tends to the adiabatic limit $\delta_{y}^2/\alpha\gg 1$, the projection probabilities should also keep constant, while it is not in this limit here because $\delta_{y}^2/\alpha\sim 0.34$. The fluctuations in the projection probabilities in the figure is due to the Landau-zener tunneling for non-adiabatic evolution. What we want to emphasize is that, although the energy gap closes and a $\mathcal{PT}$-symmetry-breaking bubble emerges when the imaginary perturbation is tuned on and is large enough, i.e., $\gamma>\delta_{y}$ [see the inset of (f); note that the gap is still open when $\gamma<\delta_{y}$], owing to the $\mathcal{PT}$ symmetry is preserved in both $H_{0}$ and $H$, the projection probabilities do not show the equality like the case with gap closing like (d). This observation has not been reported previously, since only the cases without $\sigma_{y}$ term has been discussed~\cite{longstaff2019nonadiabatic}. Therefore, the equality of asymptotic instantaneous projection probabilities only can be observed in the case where $H_{0}$ has a gap-closing transition when the dissipative perturbation is turned on, and can be employed as a unique signature of gap closing in $H_{0}$.

Although, the analysis in Ref.~\cite{longstaff2019nonadiabatic} shows that the equality of projection probabilities survive under the adiabatic condition $\gamma^2/\alpha\gg 1$ for the model $H=\eta\sigma_{z}+i\gamma\sigma_{x}$, where $\alpha$ is the tuning speed of $\eta$, the same with our model in Fig.~\ref{fig:circle_evolution} (d). In order to mimic the realistic situation, where the adiabaticity may be hard to be satisfied, we take $t_{f}=15$ and thus $\gamma^2/\alpha\sim 0.6$ here. We would like to note that, even in this case, as shown in Fig.~\ref{fig:circle_evolution}, it shows almost perfect coincidence in the projection probabilities of the final instantaneous states for the two levels. It implies this phenomenon is not strongly dependent on the adiabatic condition. Another point needs to be denoted is that, our models actually have neglected the background loss term which usually exist in experiment and thus the probabilities become larger than one. The background loss term is proportional to $\gamma$ and leads to a scaling of $\exp(-2\gamma t_{f})\sim 1/400$ in the final probabilities. It means the effective final probabilities is about $20/400=5\%$ in Fig.~\ref{fig:circle_evolution} (d). We will lose about $95\%$ of particles in the experiment.

\begin{figure*}
  \centering
  \includegraphics[width=14cm]{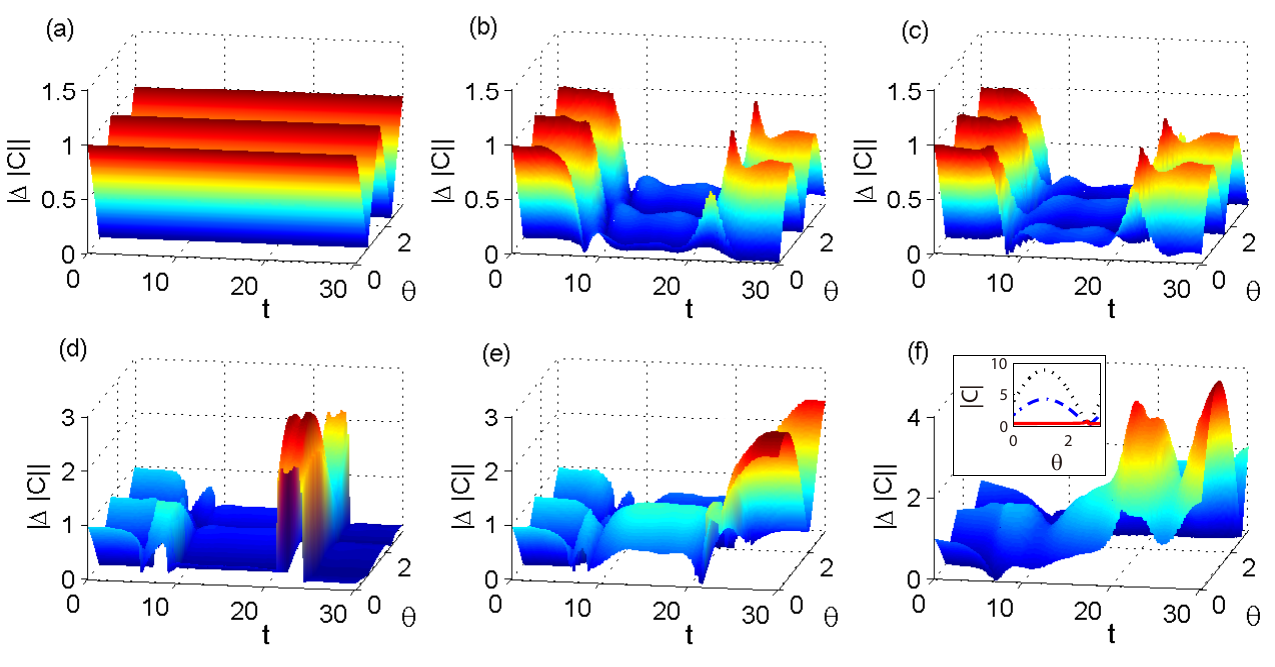}\\
  \caption{The difference between the probability amplitudes $|\Delta |C||=||C_{2}|-|C_{1}||$ for different initial states in cyclic evolution. The initial states are taken as $|\Psi(0)\rangle=\cos(\theta)|\Psi_{1}(0)\rangle+e^{i\varphi}\sin(\theta)|\Psi_{2}(0)\rangle$, where $\varphi$ is arbitrarily fixed at $\pi/6$ and $\theta$ is scanned.  These sub-figures have the same parameters as the corresponding sub-figures in Fig.~\ref{fig:circle_evolution}. The inset of (d) shows $|C_{1}|$ (blue dash-dotted line), $|C_{2}|$ (black dotted line) and their ratio $|C_{2}|/|C_{1}|$ (red solid line) at $t=t_{f}$. The ratio keeps constant except at around the zero point of $C$, which confirms the analytical result that the independence of initial states is preserved so long as the system crosses EPs. }\label{fig:amplitude_difference}
\end{figure*}

The presence of equal redistribution is also independent on the preparation of initial state. In Fig.~\ref{fig:amplitude_difference}, the differences between the instantaneous projection amplitudes are shown for different initial states. From this figure, we can find the differences between the instantaneous projection probabilities are almost absent for any initial states in the case with gap closing. In contrast, the final probability differences are not vanishing and varies for different initial states in all other cases (including the case with both anticrossing and $\mathcal{PT}$ symmetry and that with anticrossing but without $\mathcal{PT}$ symmetry) as discussed in Fig.~\ref{fig:circle_evolution}. This implies we do not need to elaborate a special initial state to identify the gap-closing transition in the experiment.

From the above discussions, the proposed scheme can be expanded into following steps. In order to identify the gap closing in a system, we first need to introduce a dissipative perturbation that is noncommutative with the original (zeroth-order) Hermitian Hamiltonian. Second, we need to tune the parameter which may lead to gap closing or anticrossing along a cyclic trajectory covering the possible critical point and observe the projection probabilities of the final states to check if the equal redistribution occurs. If the system always has a small gap, i.e., shows the anticrossing behaviour, rather than true gap closing, the equal redistribution will be broken. With these steps, we can identify the information if the system indeed has a gap-closing transition. We do not need to tune the parameter precisely to the gap-closing point, and elaborate the initial states, which are usually required in a conventional scheme. This provides us a paradigm for using dissipation in metrology.

\section{conclusion}
{In summary, we systematically characterize the nonadiabatic transitions of a generic non-Hermitian $\mathcal{PT}$-symmetric two-level model. The time evolution crossing the EPs shows initial-state-independent asymptotic behaviors. Particularly, only when the non-dissipative Hamiltonian shows gap closing, the asymptotic probabilities of particle on the two level are the same in the slow-tuning-speed limit. The equal redistribution is absent when the non-dissipative Hamiltonian displays level anti-crossing. So long as crossing EPs, the ratio between the asymptotic probabilities is initial-state independent. We thus further propose to identify gap closing with dissipative dynamics in Hermitian systems. Our proposal should be able to be checked with techniques shown in current experiments~\cite{li2019observation, wu2019observation, ren2022chiral, naghiloo2019quantum}, as exemplified in Appendix B. For example, the extension of the dynamical evolution encircling EPs in atomic gases to that crossing EPs is possible~\cite{ren2022chiral}. It is also worthwhile to note that, the idea presented here is also essentially different from the proposal to identify exceptional points with quench dynamics~\cite{agarwal2022detecting}.}

\emph{Acknowledgements}.--The authors wish to thank Profs. Wei Yi, Zhen-Biao Yang and Shi-Biao Zheng for very helpful discussions (also for the suggestions of Prof. Wei Yi on writing). J.-S. P. is supported by the National Natural Science Foundation of China (Grant No. 11904228) and the Science Specialty Program of Sichuan University (Grand No. 2020SCUNL210), F. W is supported by National Youth Science Foundation of China (CN) (Grand No. 12204105),
Educational Research Project for Young and Middle-Aged Teachers of Fujian Province
(CN) (Grand No. JAT210041) and
Natural Science Foundation of Fujian Province (CN) (Grand No. 2022J05116).

\appendix
\section{Dissipative perturbation}
We present the perturbation analysis of the impact of dissipative perturbation on a gap-closing transition or level anti-crossing in this appendix. Gap closing or anticrossing typically involves two states, the ground state $|\Psi_{g}\rangle$ and the excited state $|\Psi_{e}\rangle$, which approach and leave away each other when tuning a coupling strength $\eta$. We assume the original Hamitonian $H_{0}(\eta)$ is perturbed by a dissipative perturbation $H_{p}=\text{i}\lambda V_{p}$ with Hermitian Hamiltonian $V_{p}$ and small parameter $\lambda$, which generally can be implemented with state-dependent loss in experiment~\cite{li2019observation,ren2022chiral}. We assume $H_{0}$, $H_{p}$ and then total Hamiltonian $H=H_{0}+H_{p}$ possess the product of parity ($\mathcal{P}$) and time-reversal ($\mathcal{T}$) symmetries, i.e., the so-called $\mathcal{PT}$ symmetry: $\mathcal{PT}H_{0}(\mathcal{PT})^{-1}=H_{0}$, $\mathcal{PT}H_{p}(\mathcal{PT})^{-1}=H_{p}$ and $\mathcal{PT}H(\mathcal{PT})^{-1}=H$.  Without loss of generality, we also assume $H_{0}$ is not commutating with $H_{p}$ and thus $H_{p}$ perturbatively mix the eigenstates of $H_{0}$.

Let us employ perturbation theory to analyze the impact of $H_{p}$ on the spectra of $H_{0}$. Specifically, for the zero-order eigenstates $|\Psi_{\xi}\rangle$ and eigenvalues $E_{\xi}$, which satisfy $H_{0}|\Psi_{\xi}\rangle=E_{\xi}|\Psi_{\xi}\rangle$, the perturbation expansion of Schrodinger equation is given by
\begin{equation}\label{eq:perturbation_theory}
\begin{split}
&(H_{0}+\text{i}\lambda V_{p})(|\Psi_{\xi}\rangle+|\Psi_{\xi}^{(1)}\rangle+\cdots)=\\
&(E_{\lambda}+E_{\xi}^{(1)}+\cdots)(|\Psi_{\xi}\rangle+|\Psi_{\xi}^{(1)}\rangle+\cdots),
\end{split}
\end{equation}
where $|\Psi_{\xi}^{(n)}\rangle$ and $E_{\xi}^{(n)}$ are the $n$-th order corrections of the eigenstates and eigenenergies. It's worth noting that our unperturbed Hamiltonian is Hermitian. It is essentially different from the perturbation theory of non-Hermitian Hamiltonian~\cite{sternheim1972non}, where the zero-order basis is defined in the framework of biorthogonal theory \cite{brody2013biorthogonal}. By matching order by order, we derive
\begin{equation}\label{eq:perturbation_expansion}
\begin{split}
&H_{0}|\Psi_{\xi}\rangle=E_{\xi}|\Psi_{\xi}\rangle,\\
&H_{0}|\Psi_{\xi}^{(1)}\rangle+\text{i}\lambda V_{p}|\Psi_{\xi}\rangle= E_{\xi}|\Psi_{\xi}^{(1)}\rangle+E_{\xi}^{(1)}|\Psi_{\xi}\rangle,\\
&\cdots
\end{split}
\end{equation}
Multiplying $\langle\Psi_{\xi}|$ from left, we derive $E_{\xi}^{(1)}=\text{i}\lambda \langle\Psi_{\xi}|V_{p}|\Psi_{\xi}\rangle$, as $\langle\Psi_{\xi}|\Psi_{\xi}^{(1)}\rangle=0$.

By noting that $\mathcal{PT}iV_{p}(\mathcal{PT})^{-1}=-i\mathcal{PT}V_{p}(\mathcal{PT})^{-1}=iV_{p}$, then we derive $\mathcal{PT}V_{p}(\mathcal{PT})^{-1}=-V_{p}$. On the other hand, $\mathcal{PT}H_{0}(\mathcal{PT})^{-1}=H_{0}$, and thus $\mathcal{PT}H_{0}|\Psi_{\xi}\rangle=H_{0}(\mathcal{PT}|\Psi_{\xi}\rangle=E(\mathcal{PT}|\Psi_{\xi}\rangle)$. It follows that $\mathcal{PT}|\Psi_{\xi}\rangle=e^{i\phi}|\Psi_{\xi}\rangle$ with certain phases $\phi$, and $E_{\xi}^{(1)}=\text{i}\lambda \langle\Psi_{\xi}|(\mathcal{PT})^{-1}(\mathcal{PT})V_{p}(\mathcal{PT})^{-1}(\mathcal{PT})|\Psi_{\xi}\rangle=-E_{\xi}^{(1)}$, provided that $|\Psi_{\xi}\rangle$ is not degenerate. It implies $E_{\xi}^{(1)}=0$.

The second-order corrections of energy, $E_{\xi}^{(2)}=-\lambda^2\sum_{\bar{\xi}\neq\xi}|\langle\Psi_{\xi}|V_{p}|\Psi_{\bar{\xi}}\rangle|^2/(E_{\xi}-E_{\bar{\xi}})$, thus become dominant. We assume that the energy gap between the two states $|\Psi_{\xi=e,g}\rangle$ becomes small (anti-crossing) and even close when tuning $\eta$. In the regime where the energy gap is small, we can assume that other levels are relatively far from $|\Psi_{g}\rangle$ and $|\Psi_{e}\rangle$. The above expression of $E_{\xi}^{(2)}$ implies that the dissipative perturbation leads to the higher eigenvalue $E_{e}$ tend to decrease and the lower one $E_{g}$ increases. Then an energy gap tends to be closed by a dissipative perturbation, in contrast to the case of a Hermitian perturbation, which usually opens an energy gap or enhances the anti-crossing effect.

Assuming the two levels whose energy gap is to be considered are $|\Psi_{g}\rangle$ and $|\Psi_{e}\rangle$ with $E_{e}>E_{g}$, when the perturbation strength $\lambda\ll \Delta E\equiv(E_{e}-E_{g})$, the energy gap becomes small, but is still real and finite. The spectra are still real and the $\mathcal{PT}$ symmetry is unbroken. When $\Delta E\ll\lambda$, the dissipative perturbation becomes dominant, and degenerate perturbation can be approximately applicable. Since the perturbation matrix,
\begin{equation}
M=i\lambda\left(
  \begin{array}{cc}
    \langle\Psi_{e}| V_{p}|\Psi_{e}\rangle & \langle\Psi_{e}| V_{p}|\Psi_{g}\rangle \\
    \langle\Psi_{g}| V_{p}|\Psi_{e}\rangle & \langle\Psi_{g}| V_{p}|\Psi_{g}\rangle \\
  \end{array}
\right),
\end{equation}
is a skew-Hermitian matrix satisfying $M^{\dagger}=-M$ and thus has purely imaginary spectra. The $\mathcal{PT}$ symmetry is broken in this regime.

Although the critical value of $\lambda$ corresponding to the EPs that connect the broken and unbroken $\mathcal{PT}$ symmetry regimes cannot be fixed in the perturbation analysis, the form of $E_{\xi}^{(2)}$ indicates that it should occur when $\lambda$ is comparable to the zero-order energy gap $\Delta E$. Since $\lambda$ is small, the breaking of $\mathcal{PT}$ symmetry will only happen in the small gap regime of $H_{0}$. Therefore, a bubble breaking the symmetry $\mathcal{PT}$, with imaginary spectra inside and two EPs at the ends, will emerge when the coupling strength $\eta$ is tuned across the gaps or anticross points with a small minimum gap of $H_{0}$.

\section{Experimental Implementation}
{The analogues of a generic $\mathcal{PT}$-symmetric model $H=\eta \sigma_{z}+\delta_{0}\sigma_{0}+\delta_{y}\sigma_{y}+i\gamma\sigma_{x}$ with spin-dependent dissipations have been realized on different experimental platforms, including optical systems \cite{HH,ZPL,KKW}, ultra-cold atoms \cite{YLF,PP,JL}, ion traps \cite{ren2022chiral,WCW}, nitrogen-vacancy centers \cite{LYD,wu2019observation,liu2021dynamically} and superconducting circuits \cite{WGZ,SD,MN,PRH}. For example, for superconducting circuits, such a model can be specifically realized with a dissipative qubit with a dissipation rate of $\kappa_q$. The dissipation process can be realized by coupling the qubit with a lossy resonator with a photonic decaying rate $\kappa_r$. When no photon is leaked into the environment, the system's evolution is governed by the non-Hermitian Hamilotnian (setting $\hbar =1$)
\begin{equation}
	{\cal H}_{NH}=\Omega  (e^{i\theta}a^{\dagger }\left\vert g\right\rangle \left\langle
	e\right\vert +e^{-i\theta}a\left\vert e\right\rangle \left\langle g\right\vert )-\frac{i%
	}{2} \kappa_r a^{\dagger }a-\frac{i}{2} \kappa_q \left\vert e\right\rangle\left\langle
e\right\vert,
\end{equation}
where $\left\vert e\right\rangle $ ($\left\vert g\right\rangle $) denotes the upper (lower) level of the dissipative qubit, $a^{\dagger }$\ ($a$) denotes the creation (annihilation) operator for the resonator modes, $\Omega $ denotes the coupling strength and $\theta$ denotes the phase angle of the driving field, respectively. To modulate the coupling strength $\Omega $ and the phase angle $\theta$ in a preset sideband, an ac flux is applied to the qubit. The energy gap of the qubit is tuned in the way $\omega _{q}=\omega _{0}+\varepsilon \cos (\nu t) $, where $\omega _{0}$ is the mean $\left\vert e\right\rangle $-$\left\vert g\right\rangle $ energy difference, $\varepsilon $ and $\nu $ denote the modulating amplitude and frequency, respectively. This property enables the system dynamics to be restricted within the reduced Hilbert subspace $\{\left\vert e,n-1\right\rangle ,\left\vert g,n\right\rangle \}$ ($n\geq 1$) when the system initially has a definite quantum number, where the number in each ket denotes the photon number of the resonator. In such a subspace, we can redefine the basis vectors of the system as $\left\vert 1\right\rangle=\left\vert e,n-1\right\rangle, \left\vert 0\right\rangle=\left\vert g,n\right\rangle$. When we focus on the single-excitation case ($n=1$), the Hamiltonian of the system can be rewritten as
\begin{equation}
\begin{split}
		{\cal H}_{NH}=& \Omega (e^{i\theta}\left\vert 0\right\rangle \left\langle 1\right\vert+e^{-i\theta}\left\vert 1\right\rangle \left\langle
		0\right\vert)-\frac{i}{2} \kappa_r \left\vert 0\right\rangle\left\langle0\right\vert-\frac{i}{2} \kappa_q \left\vert 1\right\rangle\left\langle
		1\right\vert\\
		=& \Omega \cos(\theta)(\left\vert 0\right\rangle \left\langle 1\right\vert+\left\vert 1\right\rangle \left\langle
		0\right\vert)\\
        &-\Omega \sin(\theta)(-i\left\vert 0\right\rangle \left\langle
		1\right\vert+i\left\vert 1\right\rangle \left\langle 0\right\vert)\\
		&-\frac{i(\kappa_r+ \kappa_q)}{4} (\left\vert 0\right\rangle\left\langle
	0\right\vert+\left\vert 1\right\rangle\left\langle 1\right\vert)\\
   &-\frac{i(\kappa_r- \kappa_q)}{4} (\left\vert 0\right\rangle\left\langle
	0\right\vert-\left\vert 1\right\rangle\left\langle 1\right\vert).
\end{split}
\end{equation}
After rewriting it in the matrix form, we yield
\begin{equation}
\begin{split}
	{\cal H}_{NH}=&\Omega cos(\theta)\sigma_x
	-\Omega sin(\theta)\sigma_y-\frac{i
		(\kappa_r+ \kappa_q)}{4}\sigma_{0}\\
&-\frac{i(\kappa_r- \kappa_q)}{4}\sigma_z,
\end{split}
\end{equation}
where $\sigma_x=\left\vert 0\right\rangle \left\langle 1\right\vert+\left\vert 1\right\rangle \left\langle 0\right\vert$, $\sigma_y=-i\left\vert 0\right\rangle \left\langle 1\right\vert+i\left\vert 1\right\rangle \left\langle 0\right\vert$, $\sigma_{0}=\left\vert 0\right\rangle\left\langle 0\right\vert+\left\vert 1\right\rangle\left\langle 1\right\vert$ and $\sigma_z=\left\vert 0\right\rangle\left\langle 0\right\vert-\left\vert 1\right\rangle\left\langle 1\right\vert$, respectively. This is an analogue of the generic $\mathcal{PT}$-symmetric Hamiltonian that we focus on in this work. Adjusting the coupling strength $\Omega$ and the phase angle $\theta$, the corresponding nonadiabatic transitions should be observed in the experiment.
}

\bibliographystyle{apsrev4-1}
\bibliography{PT_bub_ref}
\end{document}